\documentclass[a4paper,11pt]{article}
\pdfoutput=1 % if your are submitting a pdflatex (i.e. if you have
\usepackage{jinstpub} 
\usepackage[version=4]{mhchem} 

\usepackage{gensymb}
\usepackage{textcomp}

\title{\boldmath Scintillating properties of today available lead tungstate crystals}

\author[a]{M. Follin}
\author[a,b]{, V. Sharyy}
\author[a]{, J-P. Bard}
\author[c]{, M. Korzhik}
\author[a,b]{, and D. Yvon}

\affiliation[a]{Univ. Paris-Saclay, CEA, Institut de Recherche sur les lois Fondamentales de l'Univers \\ 91191, Gif-sur-Yvette, France }
\affiliation[b]{Univ. Paris-Saclay, UMR BioMaps, SHFJ, 4 place du général Leclerc, \\ 91401, Orsay France}
\affiliation[c]{Research Institute for Nuclear Problems, Belarusian State Univ., Minsk, Belarus }

\emailAdd{megane.follin@cea.fr}
\emailAdd{dominique.yvon@cea.fr}

\abstract{
	In the context of the ClearMind project, we measured the scintillating properties, as induced from from gamma ray interactions, of today available PbWO$_4$ crystal. We measured scintillation's yields and time constants by measuring the signal shape measured on a
fast photo-multiplier and deconvoluting it from the instrumental effects. For the doped crystals at room temperature, we measured a fast scintillation component, with time constants of 2~ns, 55 \% of the total light yield, and a slow component of 6~ns. We observe a significant increase of the light yield for the slow component
when the temperature decreases and simultaneous increase of the time constants, but no increase in the fast component light yield.
Our measurements reproduce the main qualitative features of PbWO$_4$ crystals quoted in the literature. Quantitatively though, we measured significantly shorter time constants and larger light yields. This is explained by a rigorous treatment of the instrumental contributions in the measurements. Results are discussed and prospect for future developments "tailored" for the ClearMind project are presented.}


\begin{document}

\maketitle
\flushbottom

\section{Context}
\subsection{Motivations}
Positron emission tomography (PET) is a nuclear imaging technology widely used to quantify biological processes in tissues and organs by measuring the concentration of a radiopharmaceutic product \cite{Lecomte2009}. The use of PET imaging can not be widespread because of the significant dose received by the
patient, yet needed to guarantee a good image quality. The PET radio-pharmaceutical is labeled with positron emitting radio-isotopes. Positron annihilation produces two 511 keV photons emitted in quasi opposite direction. When detected, they define the so called line of response (LOR). The time-of-flight technique (TOF) uses the accurate measurement of the detection time of the two photons. It allows to enhance images contrast or reduce tracers doses delivered to the patients \cite{Conti_2019,Karp2008Jan}. Currently a commercial
system reaches a Coincidence Resolving Time (CRT) of 210 ps, and in the labs, a resolution better than 100 ps is reached with two detection modules but with a low efficiency.
Scintillator detection chain timing performances \cite{Lecoq2017} is today mainly limited by crystal scintillating properties and photo-detector performances. The development of fast photo-detector such as SiPM or MCP-PMT  and the improvement of crystals doping technologies make the "10 ps challenge" \cite{10psChallenge, 10psChallenge-2020} relevant. Such a spectacular timing accuracy would provide a 1.5 mm spatial resolution along the LORs thus making possible to get rid of back-projection and iterative algorithms and improve reconstructed images contrast a lot \cite{Budinger1983}.That would open the PET scan imaging to radiation-sensitive patient (pregnant women and children) by reducing the delivered dose (currently about 5-7 mSv for a whole-body PET) and the scan duration (currently more than 10 min).

\subsection{Detector goals}
ClearMind project is a continuation of the fast Cherenkov detectors developments CaLIPSO and PECHE \cite{Yvon2014, Ramos_2016, Canot2017, Canot2019Dec}. ClearMind aims to develop a TOF PET detection module providing tens ps CRT and high detection efficiency. It foresees high performances thanks to the detection of both Cherenkov and scintillating photons generated inside the fast monolithic scintillating crystal. Using PbWO$_4$, we expect a spatial resolution down to few $mm^{3}$.

\subsection{Short state of the art of PET detector}
\paragraph{} PET detectors can be classified according to the interaction medium which can be a crystal, a semi-conductor, a gaz or even a liquid. This paragraph focusses on inorganic scintillator properties.

As inorganic scintillators, NaI:Tl and BGO were the first crystals widely used for PET imaging. But because of the low time constant or the low crystal density, faster and/or heavier inorganic scintillators have emerged such as LSO, LYSO and LaBr$_3$, see table \ref{TabInorgaCryst}.

\begin{table}[htbp]
\centering
\caption{\label{TabInorgaCryst} Scintillating properties of principal inorganic crystals used in PET imaging \cite{Lecomte2009}. With $\rho$ for density, Phot. Eff. for photoelectric efficiency, LY for light yield and $\tau_{sc}$ for scintillation decay time. We added PbWO$_4$  for comparison,  with scintillation parameters extracted from litterature.}
\smallskip
\begin{tabular}{|l|c|c|c|c|}
\hline
Crystal & $\rho$ ($g/cm^{3})$  & Phot. Eff. (\%) & LY (ph/MeV) & $\tau_{sc}$ (ns)\\
\hline
NaI:Tl & 3.67 & 18 & 41,000 & 230  \\
BGO (Bi$_4$Ge$_3$O$_{12}$) & 7.13 & 44 & 9000 & 300/60  \\
LSO (Lu$_{2}$SiO$_{5}$:Ce) &  7.35 & 34 & 30,000 & 40  \\
LaBr$_{3}$:Ce  &   5.3 & 14 & 60,000 & 16\\
YAP:Ce  &   5.5 & 4.4 & 17,000 & 30\\
\hline
PbWO$_4$ & 8.28 & 46 & 100-300 & 1.4-22\\
\hline
\end{tabular}
\end{table}
 
The ideal inorganic scintillator would have a high light yield such as LaBr$_{3}$, would have a high density, would be made of high atomic number atoms in order to maximize photo-electric efficiency and have a fast decay time.

In commercial systems (such as the Ingenuity TF  or the Biograph mCT TOF-PET/CT systems from Philips and Siemens company respectively \cite{Kolthammer_2014,Jakoby_2011}) these crystals are coupled to PMTs, providing a time resolution about 500 ps.
More recent developments are oriented towards Magnetic Resonance Imaging compatible PET system thanks to SiPM matrix which are fast, compact and immune to the magnetic field. They provide excellent spatial resolution thanks to their one-to-one coupling (one crystal - one SiPM cell) and a high gain about 10$^{5}$ -10$^{7}$. The last commercial achievement, using SiPMs matrix and LSO crystals, is the Biograph Vision from Siemens company \cite{vanSluis2019Jan} providing a coincidence resolving time about 210 ps.

Micro channels plate photomultipliers (MCP-PMT) are also studied for PET detectors. They provide a high gain (typically about 10$^{6}$), excellent time resolution thanks to their fast transit time spread and rise time ( TTS : 35 ps and RT : 500 ps or lower 
\cite{XP85122_DataSheet, MAPMT253}) and a significant lower dark count rate about 100 Hz/cm$^{2}$ compared to SiPM detector ( 20 -200 kHz/mm$^{2}$ ). Their large active area, up to  200 x 200 mm$^{2}$ \cite{Minot2020} allows one to use large monolithic crystals as detection medium. The spatial resolution is mainly limited to the MCP-PMT pads size and the readout type (anodes pads or transmission lines).

\subsection{ClearMind detector design choices} \label{sec:DesignChoicesSec}
ClearMind project proposes a position-sensitive detector consisting of a PbWO$_4$ scintillating crystal on which is directly deposited a bialkali photoelectric layer \cite{patent_ClearMind_2017,Yvon2020}. Lead tungstate is a very dense (8.28 $g.cm^{-3}$) inorganic scintillator crystal and a good Cherenkov radiator \cite{Annenkov2002}. It produces Cherenkov and scintillating photons as a result of the 511 keV photon conversion. The light yield remains low (few hundred photons/MeV) but very fast (few ns to few ten ns, see table \ref{TabInorgaCryst}) with wide discrepancies in the published values. 
Direct deposition of a photoelectric layer on the crystal optimizes the photon collection generated inside the crystal. Cherenkov photons should significantly improve the time-of-flight information \cite{Canot2017} and scintillating photons should provide the expected spatial resolution.

The multiplication of the generated photo-electrons will be ensured by a micro-channel plate (MCP) encapsulated together with the large size crystal (up to 60 x 60 x 20 $mm^{3}$) in a standard device structure by Photek industry \cite{MAPMT253}. The readout will be performed thanks to 32 transmission lines, connected to amplifiers at both ends in order to reduce the number of  analog and digital electronic channels. 

Moddeling our detector requires detailed understanding the PbWO$_4$ scintillator crystals. We decided to measure light yields and optical photon time constants in today available PbWO$_4$ crystals.

\section{Lead tungstate crystal}
\subsection{Crystal developpment}
Lead tungstate PbWO$_4$  scintillator is a result of a development driven by a need for High Energy Physics experiments in the frame of the Large Hadron Collider (LHC) program at CERN ~\cite{Barishevsky1992, Lecoq1995}. Crystal calorimetry played a crucial role in the discovery of the Higgs boson via H-->$\gamma\gamma$  mode \cite{CMS2012}. Currently, more than 100 000 scintillation crystals are in service at two electromagnetic calorimeters at LHC: ALICE and CMS Experiments \cite{CMS1994, ALICE1995}.
PANDA Collaboration Experiment at FAIR (GSI, Darmstadt) is the next largest PbWO$_4$ based calorimeter \cite{PANDA2008}. The physics goals of the experiment required a good energy resolution near the energy threshold of at least less than 20 MeV, so it was decided to continue optimization of the PbWO$_4$ scintillator to combine high stopping power, fast scintillation kinetics with as much as possible scintillation light yield (LY). The doping program of the crystal was reconsidered. Simultaneous doping by La and Y was optimized by Crytur Company (Czech Republic) to the level, which enhanced LY by factor two. Next gain in LY by factor three, with keeping the scintillation kinetics fast, is achieved by operation at low temperatures, such as T=-25 \celsius. These, so-called PbWO$_4$-II \cite{Borisevich2016Apr} scintillation crystals, were exploited at the current research.

\subsection{Lead tungstate scintillation}
Many papers about PbWO$_4$ scintillation mechanisms have been published highlighting its luminescence centers, defects and impurities origins \cite{VanLoo1975Feb, Korzhik1994, Annenkov2002,Qu_2002}. The doping effect on luminescence properties have also been investigated in order to increase PbWO$_4$ light yield and robustness to radiations \cite{Kobayashi_1999, Nikl_2000, Mao_2002}. Let us summarize the results and current understanting of PbWO$_4$ scintillation physics.

\subsubsection{Energy relaxation}
511 keV $\gamma$ ray interacts with a crystal, mainly  through photoelectric interaction, Compton scattering and Rayleigh diffusion. The energetic "primary" free electron ionizes the matter and produces hot electron-hole pairs. These energetic free electron and hole produce secondary electron-hole pairs until their energies drop to the electron-hole pair production threshold energy, typically about 2 to 3 times the band gap energy. Then charge carriers thermalization begins: electrons and holes loose their energy by interacting with crystal lattice excitation, the so called phonons. Free electrons migrate to the bottom of the conduction band and holes to the top of the valence band, and may combine into electron-hole bounded states called excitons. The time scale of these processes in PbWO$_4$ is few ps.

\subsubsection{Scintillation mechanism} \label{sec:PbWO4_ScintMechs}
These electronic excitations (electrons, holes, excitons) transfer their energy to luminescence centers and non-radiative traps, that compete to drain the excitations energy. 
The mechanisms of scintillation in PbWO$_4$ are documented in \cite{Annenkov2002}.
Lead tungstate is a self-activated scintillator with predominant emission of quenched polaronic states.
The main luminescence center is the regular tungstate group WO$_{4}^{2-}$ which can be considered as excitations on the host oxy-anionic complexes. The Stokes shift of the luminescence is 0.42 eV, Yang-Rice coefficient is more than 6. The luminescence time constant was measured of several ns at room temperature.
But many others center contribute to the luminescence of the crytals. These are the irregular tungstate group WO$_{3}$ (few ppm), and MoO$_{4}^{2-}$ (~ppm or less) due to residual contaminations in the crystal, Molybdenum replacing Tungsten atoms in the matrix host. Some of the WO$_{3}$ centers are located near a Frenkel defect, which impact a lot on their properties. Finally, an additional luminescence center associated to lead defect in the crystal and the consecutive crystaline structure modification has been identified.
	Worth to noting, both regular WO$_{4}^{2-}$ and defect-related WO$_{3}$ luminescence centers show the same very short (~ps) leading edge of the luminescence kinetics. This is an indication that no intermediate recapturing processes are involved in the energy transfer processes. 

\paragraph{}
	Once activated a luminescence center can relax through three competing  processes: thermal ionization to the conduction band, non radiative decay (thermal quenching) and emitting photons. Mean ionization time depends on the energy gap between the trap and the bottom of conduction band. Thermal quenching time constant depends on the luminescence Stokes shift which is found to be large for all radiation centers in PbWO$_4$. Both ionization and quenching are thermally assisted and thus depend strongly on temperature.
  WO$_{4}^{2-}$ centers emit photons in band centered at 420 nm (\emph{blue} luminescence) by radiative recombination of its polaronic states. Its trapping energy is 50 meV, thus very small. Because of thermo-ionization, these centers are very unstable at room temperature, therefore, the light yield remain small.  WO$_{3}$ centers emit photons at 480-490 nm ( \emph{green} luminescence). Their trapping energy is much larger, 0.53 eV, and they contribute significantly to crystal light yield. MoO$_{4}^{2-}$ centers were shown to radiate at 508 nm. Finally the Frenkel defect center was shown to radiate in the red spectral range. These are deep stable traps with long time constants, tens of ns decay times at room temperature.
  
  When PbWO$_4$-II crystals are doped with trivalent La and Y ions at the total level less than 100 ppm, the regular WO$_{4}^{2-}$ radiative recombination centers are distorted by nearby lanthanum La$^{3+}$ and yttrium Y$^{3+}$ ions. The WO$_{4}^{2-}$ + Y$^{3+}$ and WO$_{4}^{2-}$ + La$^{3+}$ groups are formed. They emit at 420 nm. Their thermo-activation energy is significantly enlarged (130-200 meV) compared to regular WO$_{4}^{2-}$ centers, thus these centers are stable and form a set of luminescent centers similar to doping ions in inorganic crystals. Trivalent rare earth centers were found, in addition, to improve greatly crystal radiation hardness and to prevent the deeper centers to fill, thus preventing scintillation at slow time constants.

\subsubsection{Fast luminescence of "lead tungstate II" crystals} \label{sec:PbWO4_FastLum}
\paragraph{}
In spite detailed spectroscopy of the PbWO$_4$ crystals in early days of development, only recently fast decaying luminescence kinetics components were measured in PbWO$_4$-II crystals \cite{Auffray2016Oct}. The very fast luminescence showed two ps-range decay components: 5.9 and 824 \emph{ps}, at 254 nm interbands excitation.

The ultrafast decay component was found to be due to green luminescence emitting centers WO$_{3}$. Those centers are very rare and the measured light yield is extremely small, less than 0.1\% of the total light yield. 
The medium decay component (600-800 ps), corresponds to the radiative recombination of WO$_{4}^{2-}$ + Y$^{3+}$ and WO$_{4}^{2-}$ + La$^{3+}$ groups. 
Finally the well known scintillation of regular WO$_{4}^{2-}$ centers, of several ns is also observed.

Many papers have been published during the 30 years long work. The measured light yield, decay times are shown to strongly depend on crystal manufacturing options, doping and remaining measured impurities in the components used to manufacture the crystal. Futhermore, as expected, because of the strong thermal quenching of radiative centers, measured properties are strongly dependant on temperature : the lower the temperature is, the higher the light yield is and the slower time constants are.  Reference \cite{Mao_2008} documents a light yield loss of (-2.5 $\pm$ 0.1) $\%/ \celsius$ for PbWO$_4$ scintillators. 

In order to allow accurate detector modelling and optimisation, we decided to measure the properties of today available lead tungstate crystals. Following are the methods used and our results.

\begin{figure}[htbp]
\centering % \begin{center}/\end{center} takes some additional vertical space
\includegraphics[scale=0.30]{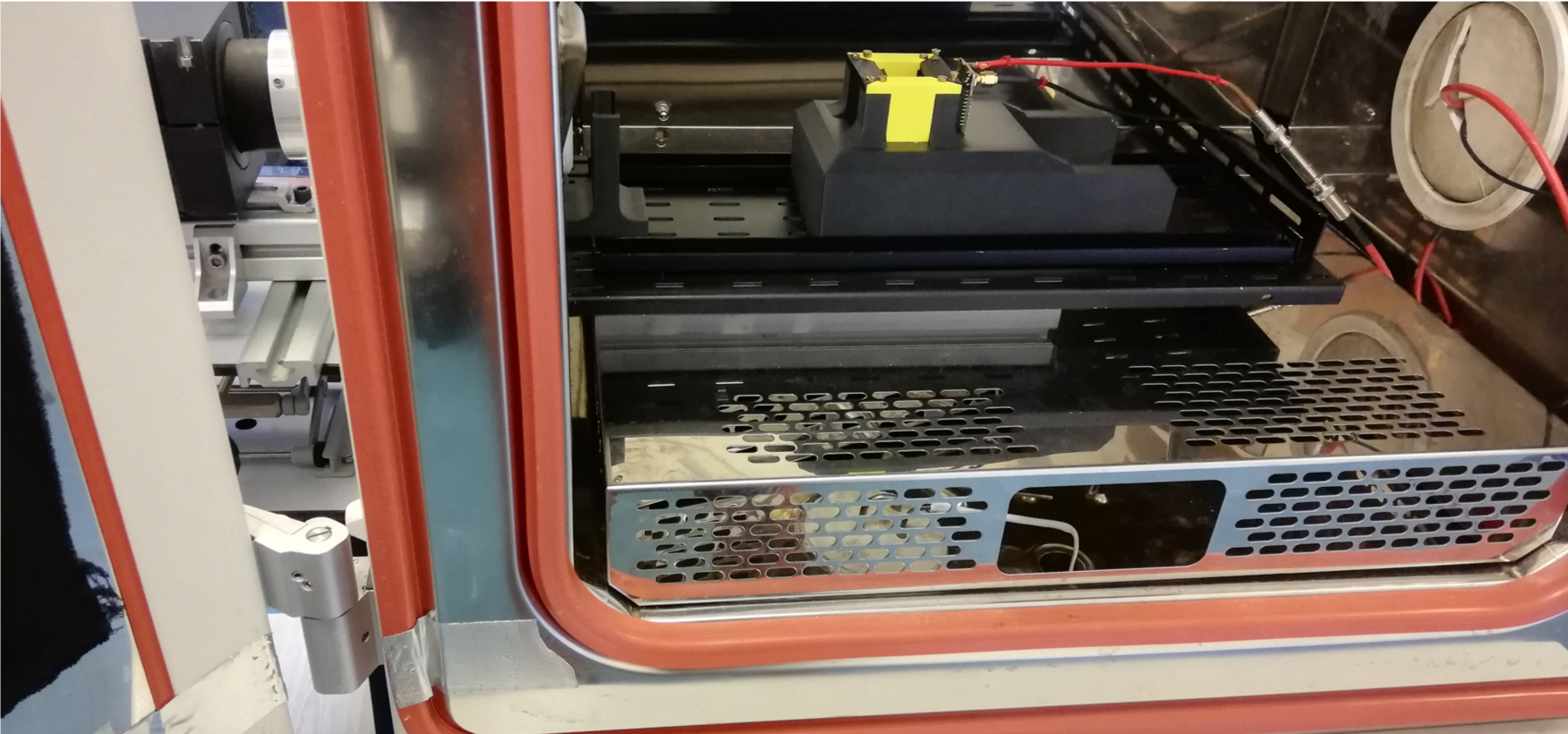}
\qquad
\includegraphics[scale= 0.35]{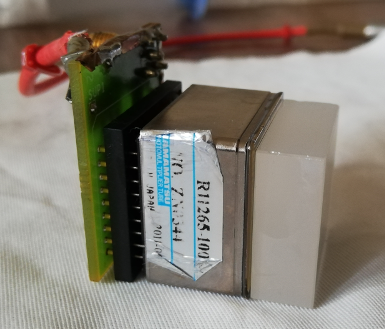}
\caption{\label{fig:CaractBench} Left : crystals bench caracterization. Right :  R1126-100 photomultiplier coupled to a PbWO$_4$ crystal thanks to optical gel.}
\end{figure}

\section{Measurement  setup and methods}
\subsection{Crystals characterization bench} \label{sec:CrystalBench}
\paragraph{} Four PbWO$_4$ crystals of size 25 x 25 x 10 mm$^3$ are studied. Three crystals are doped and were provided by CRYTUR \cite{CryturComp} (x1) and SICCAS Corp. \cite{SICCAS_Comp} (x2). CRYTUR's crystal uses Panda-II technology and crystals from SICCAS are doped with CMS doping (labelled SICCAS CMS) or with Yttrium (SICCAS Y). The fourth crystal is undoped and was provided by EPIC-Crystal \cite{EPIC}. All crystals are five sides grounded in order to maximize the probability of optical photons collection. The coupling face to the photocathode is polished and crystals are attached to the photocathode thanks to an optical gel (SmartGel OCF-452 from Nye).

The photodetector used for PbWO$_4$ crystal characterization is a photo-multiplier R11265-100 from Hamamatsu, biased using a home made tappered HT-positive resistor network. This classical setup, usually used in photon counting apparatus, allows one to ground the photocathode layer, to minimize PMT dark noise and to improve PMT assembly time resolution on single photo-electron signal. The PMT gain is about  1.4 10$^{6}$ and the TTS (\emph{Time Transit Spread}) is typically 270 ps at 25 \celsius, according to the datasheet. Its bialkaline photocathode provides a 23 mm x 23 mm active area.
The \emph{ "photodetector + crystal"} system is marked as BRS photodetector in this article.

We used a $^{22}$Na radioactive source which emits $\beta+$ particles and 1.27 MeV $\gamma$ rays. The fast YAP spectrometer \cite{Ramos2014} is used to acquire, in coincidence with the BRS photodetector, events produced from the two 511 keV photons emitted back-to-back following the electron-positron annihilation.

BRS photodetector electrical signals are amplified thanks to a ZKL-1R5+ amplifier \cite{ZKL-1R5+} with 1.5 GHz bandwidth and gain 40 dB.
The acquisition module is a WaveCatcher \cite{Breton_2020} that sample both BRS detector and YAP spectrometer pulses shapes at 3.2 GHz in a time window of 320 ns.

The BRS photodetector is powered by +750 V and the YAP spectrometer by -1200V. The \emph{"radioactive source + BRS photodetector"}  system is placed in a Vostch VC 4018 climatic chamber in order to adjust and regulate the system temperature from 20\celsius to -25\celsius. The YAP spectrometer is placed outside the climate chamber, in line with the the $^{22}$Na and BRS detector, see figure \ref{fig:CaractBench}.

\begin{figure}[htbp]
\centering % \begin{center}/\end{center} takes some additional vertical space
\includegraphics[scale= 0.45]{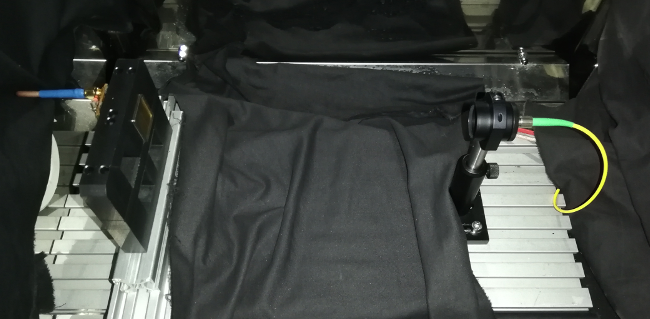}
\caption{\label{BenchCalib}Calibration bench :  PMT (left) and laser fiber on support (right).}
\end{figure}

\subsection{Photomultiplier calibration bench} \label{sec:Calib_Bench}
\paragraph{} 
The optical parts are placed into the climatic chamber in order to acquire datas in the same readout conditions as when used for crystals caracterization, see figure \ref{BenchCalib}.

We use uses a "Pilas" 409 nm laser from "Advance Laser Diode System" \cite{PilasLaser2020} for the light source. The laser output connect to the test apparatus using an optical fiber. In these conditions the optical beam time distribution is documented to be gaussian-like with 20 ps FWHM and the jitter on the trigger signal is 1.4 ps. The laser fiber is placed at a distance of 28 cm from the photomultiplier, directed towards the photomultiplier. We  attenuated the beam so that the BRS photomultiplier triggers in \emph{"single photon electron"} mode. The laser head and its power module are placed as far as possible (i.e. 4 m) of the test bench and readout electronics in order to minimize the induced electronic noise.
BRS photomultiplier is readout with the same electronic as documented in section \ref{sec:CrystalBench}. The Wavecatcher module triggers on a coincidence between the BRS Photomultiplier and the Pilas laser,  and registers the BRS and the Pilas laser synchronisation pulseshape. 
A calibration run is acquired before each measurement cycle. Typically we registered 100000 events triggered on signals (threshold 1/3 of mean Photo-electron amplitude).

\begin{figure}[htbp]
\centering % \begin{center}/\end{center} takes some additional vertical space
\includegraphics[scale= 0.50]{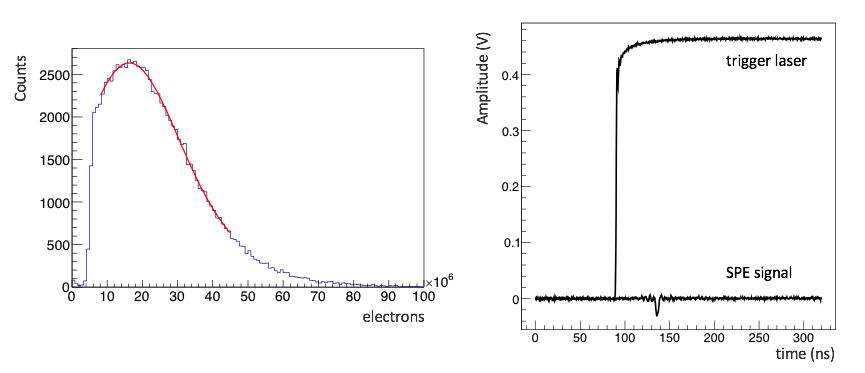}
\caption{\label{IntegChargeSPE} Left : histogram of single photo-electron charge  after the gain 100 amplifier. The red line represents the fit to the charge distribution. Right: laser trigger and Single PE timeline. }
\end{figure}

\subsubsection{Signal processing} \label{sec:SignalProcess}%\subsection{Calibration results}
\paragraph{} 
To quantify the single photo-electron signal and reduce the random noise, we sum together the 100 000 events using the Pilas trigger signal as a time reference. The mean charge integral labeled as IntegCharge$_{SPE}$ is calculated over a 10 ns time window (T$_{Integ}$). A \emph{mean shape}  of single photoelectron signals is also computed MeanShape$_{SPE}$(t).
We checked and found that the results from calibration runs do not depend on system temperature.
We noticed that the mean SPE charge does drift after being powered. This is likely to be due to a gain versus temperature dependence of the fast post amplifier used in the readout. The measurement was found to be stable after 45 min.

\begin{figure}[htbp]
\centering % \begin{center}/\end{center} takes some additional vertical space
\includegraphics[scale= 0.60]{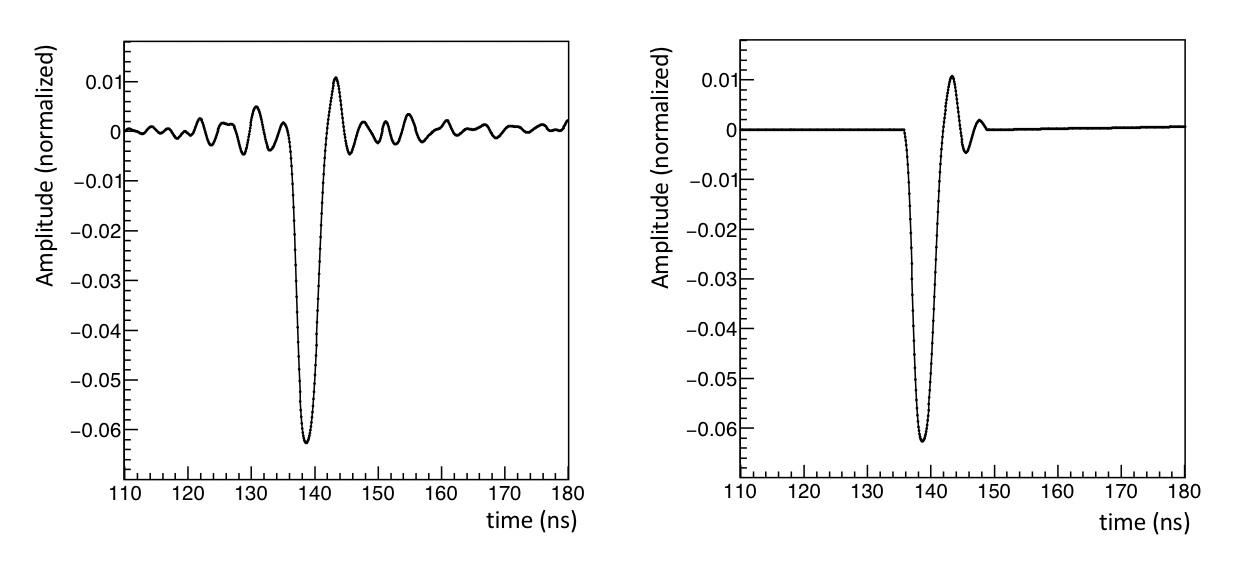}

\caption{\label{fig:MS_SPE} Single photo-electrons mean shape. Left: Raw mean shape as averaged from 100000 pulses. Optimized single photo-electrons pulse shape, after careful filtering of time synchronous parasitics induced by the Pilas laser (see section \ref{sec:SignalProcess})}.
\end{figure}

We computed the pulse time as the time interpolated at 50\% amplitude on pulse rising edge (digital Constant Fraction Discriminator algorithm).
The histogram of the collected charge, is adjusted the sum two gaussians (see Figure \ref{IntegChargeSPE}). We compute, after the amplifier, a mean charge integral value of the photoelectron of $(1.61\pm0.08) 10^{7}$ electrons.
This step requires some care. The signals shape was oversampled by the factor 4 in order to minimize the distortion related to the bin size during the synhronisation with laser trigger signal. It is marked as MeanShape$_{SPE}$(t) as mentionned in section 3.4.1 and figure \ref{fig:MS_SPE}.

\subsubsection{Optimizing single photo-electron signal mean shape} \label{sec:SPE MeanShape}
\paragraph{} 
	In order to measure the SPE pulse shape, we needed a very fast pulsed light source: the Pilas picosecond laser. Its light pulse width of 20 ps match our need. But the laser head induces very fast time synchronous parasitics on the BRS readout that bias single photo-electron pulse shape and, in the end, the measurements. Though we averaged tens of thousand pulses, the MeanShape$_{SPE}$(t) showed to be noisy before the pulse (so called baseline), and after pulse. Over time we improved the setup. We shielded the temperature controlled dark box used for the measurement, changed the BRS readout amplifier, used a 5 m long optical fiber between the Pilas laser output and the measurement setup. Those modifications reduced significantly this noise.	Nevertheless the picosecond laser electronics still induced significant parasitics on BRS signals, synchronous with pulses, thus not averaging out. This noise has to be minimized without filtering the main pulse shape.
	
	As no physical signal is expected in the  pre-pulse zone, we decided to set the pulse shape value at 0 in this zone. 
	The post-pulse zone (defined as the zone after the SPE pulse has decayed) is more complex. The main pulse shape involves a fast decaying oscillating behavior that we have to keep. In addition, choosing the high voltage positive biasing configuration for the PMT induces a pulse overshoot slowly decaying with a 500 ns time constant. A choice had to be made when to start filtering the later pulse shape. We tested several options and selected the one minimizing the $\chi^{2}$ when fitting measured PbWO$_4$ light pulses. We kept three following "ringing" oscillations after the main pulse and then filtered the later pulse shape thanks to a non parametric gaussian-kernel function, centered in each point of the region of interest (SmoothKern of ROOT (CERN) library, Filter BW 15). The resulting "optimized" SPE mean shape is labeled in the following as MeanShape$_{{SPE}_{Opti}}$(t).
	Both raw and optimized SPE mean shapes are shown in figure \ref{fig:MS_SPE}.

\begin{figure}[htbp]
\centering 
\includegraphics[scale= 0.65]{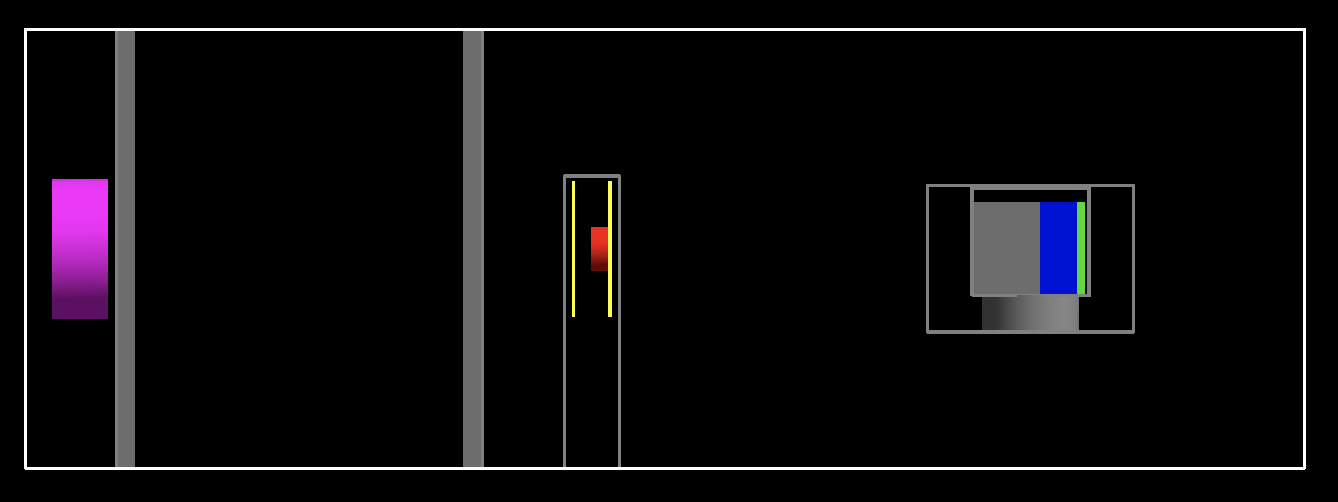}
\caption{\label{fig:MOC} Simulated crystal caracterization bench via GATE. The YAP crystal is in magenta and the radioactive source is in red inside the yellow plastic support. The two grey bands (in aluminum) represent the climatic chamber walls. The lead tungstate crystal is in blue, the window of the photomultiplier is in green and the photocathode in red (too thin to be visible here).}
\end{figure}

\subsection{Light collection model : a GATE simulation } \label{sec:LightCollModel}
\paragraph{} Modelling our measured BRS pulse shapes requires to understand light collection in the BRS detector. Assessing photon light yield from the measured photoelectrons statistics requires understanding light collection and photoelectron production efficiency. In order to do so, we used a Mont\'e-Carlo simulation running under the GATE open source software \cite{GATE2004} driving the GEANT4 simulation package \cite{Agostinelli_2003, Allison2016Nov}.
A realistic (though simplified) geometry of the measurements has been computed (see figure \ref{fig:MOC}). The simulated source emits two 511 keV gammas back-to-back. Physics models used for the PbWO$_4$ crystal and bialkali photocathode are documented in \cite{Yvon2020}. The BRS PMT glass window and optical gel have been modeled as a 2~mm thick BK7 glass. This allows a fair simulation of the optical reflections and refractions at the crystal/optical gel boundary.

\begin{figure}[htbp]
\centering 
\includegraphics[scale= 0.55]{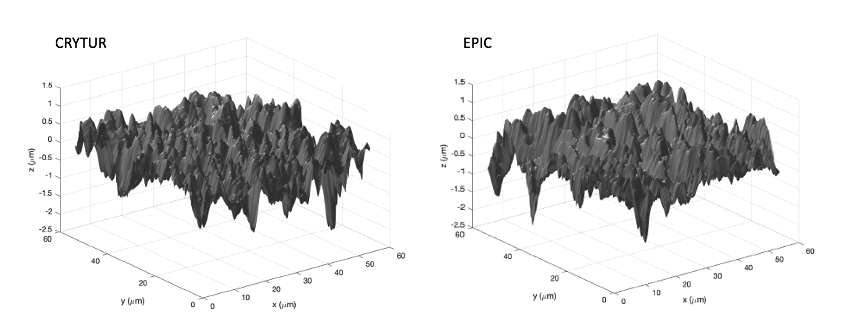}
\caption{\label{fig:MesCrysSurf} Maps of surfaces depths, of two of our PbWO$_4$ samples, as measured on a Keyence VK-X100 confocal microscope.}
\end{figure}

\subsubsection{Simulation of grounded crystal optical surfaces and photon collection time distribution}\label{sec:SimuOptSurf}
\paragraph{}
The PbWO$_4$ crystal refraction indexes being very high (from 2.2 to 2.4 at optical wavelengths), when a photon impinges at crystal/air interface, it is very likely to be reflected in total reflection mode. This is also true to a lesser extend at the crystal/optical gel interface. We decided to ground the crystal to air surfaces in order to randomize the refection angle. This allows optical photons propagating to impinge the crystal/optical gel interface at varying angle thus increasing their probability to go through and to be collected by the photocathode.

The grounded crystal surfaces have first been simulated using the unified model \cite{Levin1996} labeled as "rough grounded" with a scattering parameter $\sigma_{alpha}$ to be chosen by the user.
We simulated  $10^5$ events, assuming light production through Cherenkov and scintillation with 300 photons/MeV scintillation light yield and a decay time of \emph{1 ps}. 
We observed that different values for the $\sigma_{alpha}$ parameter, i.e. 6, 20 (as measured by \cite{Roncali2013Mar} on a LYSO crystal) or $50^{\circ}$, led to very different light collection time distribution, that impact significantly on the recovered scintillation properties. We thus needed to quantify our crystal roughness and diffusion properties.

We used a Keyence VK-X100 confocal microscope configured in "expert, top surface" mode from the Orsay Panama measurement platform in order to image the surface depth of our four PbWO$_4$ samples. We measured surface depth maps shown at figure \ref{fig:MesCrysSurf}, with rugosities ranging from  1.29 $\mu m$, to 1.43 $\mu m$ (quadratic rugosity Rq).
The measured data were converted into Look Up Table (\textbf{LUT}) with the dedicated software of UC Davis team that computes the optical scattering probability from a measured surface depth [44, 45]. These LUTs have been interfaced with Geant4 and Gate simulation packages and used in our Monté Carlo simulation. We observed that the light collection time distribution
of two crystals of smallest and largest rugosity are very similar and that using one or the other in the following analysis does not impact the recovered scintillation parameters.

\begin{figure}[htbp]
\centering 
% \begin{center}/\end{center} takes some additional vertical space
\includegraphics[scale= 0.4]{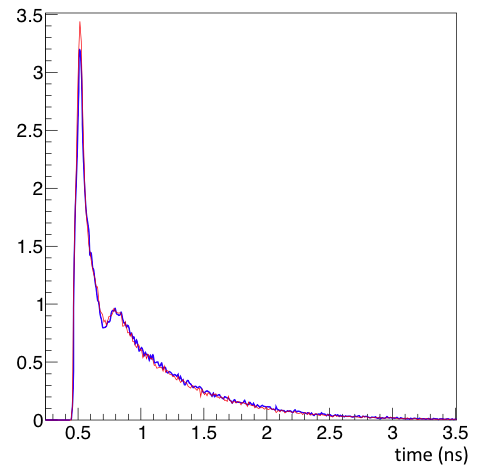}
\caption{\label{fig:PhotCollec} Simulated time distribution of photon collection by the bialkaly photocathode, assuming instantaneous light production. \emph{Blue} : Time distribution computed using UC Davis Optical surface software and the rugosity map measured on the grounded CRYTUR PbWO$_4$ crystal. \emph{Red:} Time distribution computed using unified optical surface with $\sigma_{alpha}= 20^{\circ}$.}
\end{figure}

This study allowed us to compute a realistic distribution of the optical photons collection and photo-electron production times in our apparatus and in addition the light collection efficiency of the photocathode : 33\%.
Figure \ref{fig:PhotCollec} shows the simulated photon collection time distribution computed from the measured Crytur crystal rugosity map and, for reference, the time distribution computed using the unified model assuming $\sigma_{alpha}$ of $20^{\circ}$.

\subsubsection{Computing Cherenkov light photo-electron Yield}
We then ran a simulation assuming that Cherenkov light was the only optical light production mechanism, in order to compute its contribution to the detected photoelectrons yield. Given our crystal physics model (all crystals have the same optical transmittion curve), this contribution is assumed to be the same for all measured PbWO$_4$ crystals. We computed a mean yield of 0.80 photoelectron from Cherenkov light.

\subsection{BRS data analysis} \label{sec:BRSAnalysis}
\subsubsection{Light yield measurement}
For each crystal and  temperature we registered 30k events, triggered by a coincidence between YAP spectrometer (threshold 230 keV) and BRS detector (threshold at 1/3 PE). Pulse Shape are quantified in time (Constrant Fraction Discriminator algorithm) and collected charge is integrated over time both for the BRS and the YAP timelines.
We then built the histogram of the charge integrated over the BRS timelines. This histogram can then be normalised by the single photoelectron mean charge value. 
\begin{equation}
Nb_{PE_{{crystal}_{T}}} =   IntegCharge_{{crystal}_{T}} /  IntegCharge_{SPE}
\end{equation}

With IntegCharge$_{{crystal}_{T}}$, the mean charge integral measured through the crystal at T $ \celsius$.
This allowed us to reconstruct a raw estimation of the collected photoelectron number at the temperature T $\celsius$ on this crystal.

\subsubsection{Scintillation components measurements} \label{sec:ScintTimeMeas}
We then built the mean BRS signals shape, for each crystal and temperature, by averaging the BRS signals pulses using for the reference time the CFD time of the YAP spectrometer. 
The algorithms used are the same as for the single PE mean shape computation. Once again this procedure allowed us to reconstruct rise time on averaged signals minimaly distorted by the well studied YAP spectrometer time resolution. The pulse shape is then normalised to an integral of one. This mean shape is labeled as MeanShape$_{{crystal}_{T}}$(t). 

The pulse shape is a convolution of the four following contributions : \newline 
$\bullet$ the photons production and collection time distribution (511-keV gamma propagation, interaction in the crystal, the optical photon collection and detection by the PMT photocathode) \newline 
$\bullet$ the YAP spectrometer time distribution \newline 
$\bullet$ the single photo-electron mean shape (from the calibration run) \newline 
$\bullet$ and \emph{the time distribution of optical photons production} inside lead tungsten crystals, by scintillation and Cherenkov processes, so called \emph{scintillation model}\newline 

The first item is computed by the  GATE simulation as explained in section \ref{sec:LightCollModel} and the corresponding time distribution D$_{Col}$(t) is shown in figure \ref{fig:PhotCollec}.

The YAP spectrometer time resolution has already been measured to 105 ps (standard deviation) \cite{Ramos2015}. It is thus modeled by a normal law of parameter ($\mu$=0, $\sigma$=105 ps): 
\begin{equation}
D_{\mathrm{YAP}}(t) = \frac{1}{\sigma\sqrt{2\pi}} \exp \left( \frac{-t^{2}}{2\sigma^2} \right)
\end{equation}

The single photo-electron (SPE) mean shape (Figure \ref{fig:MS_SPE}) is obtained from the calibration runs see section \ref{sec:Calib_Bench}. 

Finally, the mathematical model of light production due to scintillation is defined as the sum of decreasing weighted exponential. Each exponential term describes a luminescent center kinetic  providing a specific decay time. Cherenkov light production is taken into account as very short (single sample) light component.

Follwing the literature (see section 2.3), light production inside undoped lead tungsten crystals can be modelled as : 

\begin{equation} \label{eq:ScintModel}
D_{scinti}(t)= R_{ch}\delta(t) 
	+ \frac{R_{f}} {\tau_{f}} \exp \left( \frac{-t}{\tau_{f}} \right) 
	+ \frac{R_{slow}} {\tau_{slow}} \exp \left( \frac{-t}{\tau_{slow}} \right) 
	+ \frac{R_{Vslow}}{\tau_{Vslow}} \exp \left( \frac{-t}{\tau_{Vslow}} \right) 
\end{equation}

with: R$_{ch}$ the Cherenkov light yield , R$_{f}$ the scintillation light yield associated to the fast time constant $\tau_{f}$, R$_{slow}$ the scintillation light yield associated to the slow time constant $\tau_{slow}$ and R$_{Vslow}$ the scintillation light yield associated to the very slow time constant $\tau_{Vslow}$. 

\begin{figure}[htbp] 
\centering % \begin{center}/\end{center} takes some additional vertical space
\includegraphics[scale= 0.6]{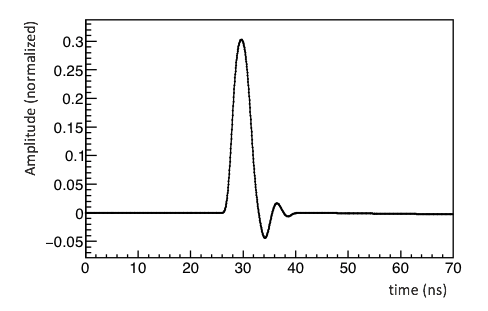}
\caption{\label{fig:DetectorShape} The Instrument shape, D$_{Instrument}$, is the shape we would measure, would light production in the crystal be instantaneous. It is the convolution of light collection distribution, YAP spectrometer time resolution and BRS pulse shape.}
\end{figure}

For doped lead tungsten crystal the same model can be used, but the slow time constant was measured to be negligible. Thus when usefull it will be set to zero in the fitting procedure.

Our modelled pulse shape f(t) is the convolution of these 4 contributions. 
\begin{equation}
f(t) =  D_{Col} (t) \circledast D_{YAP}(t) \circledast  MeanShape_{{SPE}_{Opti}}(t) \circledast  D_{scinti}(t) 
\end{equation}

The convolution pitch was chosen to 25 ps. The convolution of the three first contributions D$_{Instrument}$ the \emph{instrumental shape} (see figure \ref{fig:DetectorShape}) can be computed once for all, once the PMT $MeanShape_{{SPE}_{Opti}}$ has been calibrated. This shape corresponds to an instantaneous production of light in the crystal. 
\begin{equation} 
D_{Instrument} =  D_{Col} (t) \circledast D_{YAP}(t) \circledast  MeanShape_{{SPE}_{Opti}}(t) 
\end{equation}

\begin{figure}[htbp] 
\centering % \begin{center}/\end{center} takes some additional vertical space
\includegraphics[scale= 0.47] {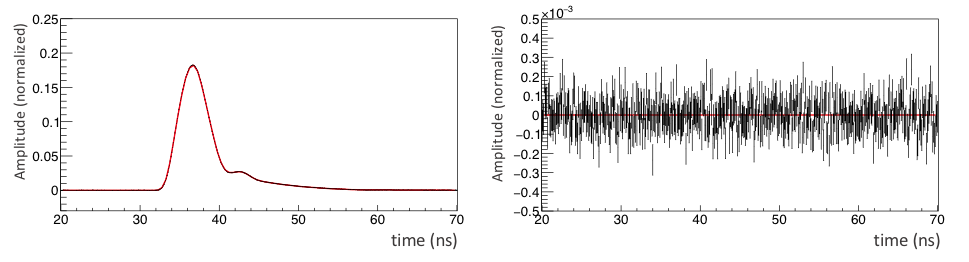}
\caption{\label{fig:SimuMeasure} 
\emph{Left:} Simulated pulse shape (Black line) according to the measured scintillation parameters of a SICCAS Y-doped crystal at 20 \celsius. We added white noise to the simulated timeline and ran the analysis software. The fittted pulse shape is shown in red. 
\emph{Right:} Difference between the simulated and the fitted pulse shape.
Pulse shape is fitted very accurately, and error timeline shows the white noise structure.
}
\end{figure}

A modelled pulse shape, parametrised by the scintillation model free parameters can thus be computed very efficiently.
We then used the fitting tools available in the ROOT analysis software package \cite{Root_Brun_1997} to adjust the scintillation model parameters in order to reproduce the averaged pulses measured on PbWO$_4$ crystals (figure \ref{fig:SimuMeasure}). This allowed us to quantify the light yields and decay times of the scintillation components in the crystal versus the temperature.

\section{Method accuracy and biases: a Monte Carlo study} \label{sec:MethodAcc}
\paragraph{} 
In order to assess and quantify the proposed method accuracy and biases, we decided to implement testing procedure using a Mont\'e Carlo method.
We simulated PbWO$_4$ pulses by convolving the averaged instrumental shape (figure \ref{fig:DetectorShape}) calibrated on data with an assumed scintillation model of equation \ref{eq:ScintModel}. We then added to this shape a random gaussian noise of width chosen to match the measured value (8~10$^{-5}$) on the prepulse timeline of averaged pulse in our experiments. 

We proceeded step by step of increasing complexity. We first input a simple two components pulse shape (Cherenkov 10\% + scintillation 2ns decay time, 90\%), then additional short and slow scintillation components  (Cherenkov 10\% + scintillation decay time 2ns, 30\%, 5ns, 40\%, 15ns, 20\%).

\subsection{Ultrafast time constant accuracy and bias}
We simulated very short scintillation time constant. We expect that reconstruction becomes difficult because ultrafast scintillation is difficult to distinguish from the Cherenkov component given the pulse risetime of our photomultiplier.
For this study, we assumed a scintillation model of very slow time constant 15 ns, 20\% yield, slow 5ns, 40\%, a Cherenkov yield of 10\% and a fast time constant varying from 2ns down to 0.4 ns, 30\% yield).
Studies shows that the scintillation parameters are reconstructed accurately for fast time constant values larger that 1 ns. At shorter scintillation time constants, reconstruction "blends" the Cherenkov and the ultrafast time constant : the reconstructed Cherenkov yield decreases, the fast component yield "collects" most of the missing Cherenkov yield, slow time constants are slightly biased toward low values. But the Cherenkov light yield can be computed using our Mont\'e Carlo simulation. Thus we decided to set the Cherenkov component yield to its theoretical value in the reconstruction. Under these conditions, the reconstruction algorithm does not show any significant (less than 1 \%) bias anymore, for fast scintillation time constant down to 0.4 ns. For all further works, the Cherenkov light yield is set to its simulated value.

\subsection{ Method accuracy assessment - statistical errors}
We finally simulated and reconstructed 200 measurements assuming scintillation parameters close to those measured at room temperature on Y-doped crystals (Siccas Y-doped and Crytur-Panda II): Cherenkov yield 5\%, fast scintillation 1.3 ns, 47\%, slow, 5.5 ns, 48\%. As expected from results above, reconstruction algorithm converge with no significant bias. Dispersion on reconstructed scintillation parameters are computed very small. We find 1 $\sigma$ values on fast, slow, scintillation yields and time constants of less than .2\%.

\begin{figure}[htbp] 
\centering % \begin{center}/\end{center} takes some additional vertical space
\includegraphics[scale= 0.50] {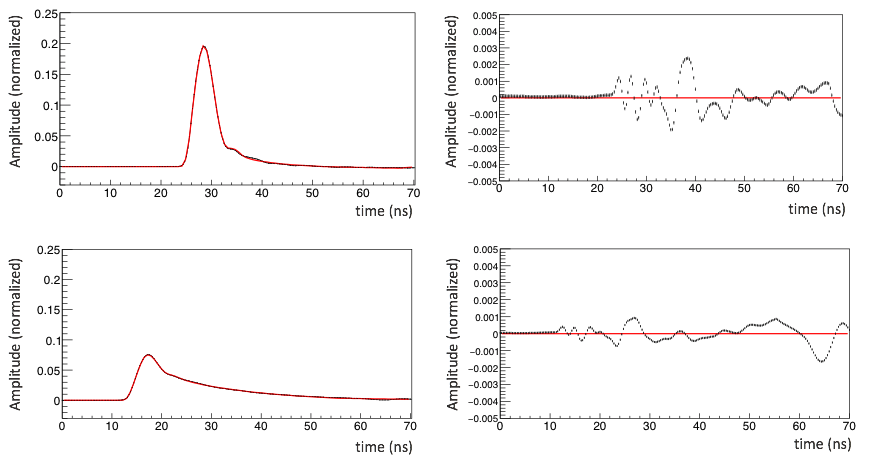}
\caption{\label{fig:Measure&Fits} 
\emph{Left:} Measured pulse shape (black lines) fitted pulse shapes (red lines) on a Crytur crystal at 20~$\celsius$ (top) and -25 $\celsius$ (bottom). 
\emph{Right}: Difference between the measured and the fitted pulse shape. Please note the difference in scale.Though the fitted line match closely the measured mean pulse shapes, we observe residual structures in the timelines of differences. See text for discussion.
}
\end{figure}

\section{Data Analysis}
\subsection{Additional Instrumental features}\label{sec:AddInstruFeatures}
\paragraph{}
Reconstruction algorithm is thus very accurate, if a very accurate, "perfect" model is available for the instrument. Unfortunately measurements turned out to be demanding given the technology available. 
Figure \ref{fig:Measure&Fits}, \emph{left} shows typical data from our measurements at 20 $\celsius$  and -25 \celsius, our highest and lowest temperatures, on a CRYTUR Panda II crystal.
The optimized mean shape shown at figure \ref{fig:MS_SPE}, improved the fit $\chi^{2}$ from one to three order of magnitude compared to using the raw averaged shape. We observe that the model fits  well the averaged pulse shapes
Figure \ref{fig:Measure&Fits}, \emph{Right} plots the \emph{difference} between the measured averaged pulses shape and the fitted shapes. Please notice the change in amplitude scale. We notice a residual difference of the order of 1\ \% of main amplitude after the main pulse. Though these differences are small, they are not structured as a white noise and are much larger that the RMS noise measured on the pre-pulse part of the timeline. Thus the reconstructed $\chi^{2}$ is large.
These residual differences are not averaged out with more statistics and we associated them with the laser-induced noise.
%Including the three damped oscillations in the SPE Shape allowed to minimize the amplitude of the residuals in the time window between 20 and 40 ns. Filtering the post pulse shape minimized significantly the residual between 40 ns and 70 ns. The largest remaining feature in the residual curve is then the bump 32 and 42 ns.
%	The residual curve ves time look very much alike when measured on runs at the same temperature on different crystals.  Assumptions on the surface diffusion model (Davis LUT or unified model, 20 \degree) do not impact on them significantly. It is thus likely that the remaining errors are due to laser-induced noise in the unfiltered part of the SPE shape. These are remaining imperfections in our instrument model that would additionnal hardware enhancement in order to improve them.

%
%
%
\begin{table}[htbp]
\centering
\caption{\label{tab:Doped_PbWO4_Results} Measured luminescence properties of doped PbWO$_4$ scintillator crystals on our apparatus for a mean energy deposited of 432 keV. $Y_{xx}$ stands for photo-electron yields, $\tau_{XX}$, for scintillation time constants. The Cherenkov contribution to yield, 0.80 photo-electron, is included in the total luminescence Yield.
Second part : Computed values of systematic errors on measurements (see paragraph \ref{sec:Systematics}) 
}
\smallskip
\begin{tabular}{|c|c|c|c|c|c|}
\hline
Temp. & $Y_{Total}$ & $Y_{Fast}$ & $\tau_{Fast}$ & $Y_{Slow}$ & $\tau_{Slow}$ \\
(\celsius) & (PE) & (PE) & (ns) & (PE) & (ns) \\
\hline
\multicolumn{6}{|c|}{CRYTUR - Panda II} \\ \hline
20 & $15.2\pm0.5$ & $8.45\pm0.1$ & $1.80\pm 0.06$ & $6.0\pm 0.3$ & $6.4\pm0.2$ \\
5 & $22.3\pm0.5$ & $8.9\pm0.1$ & $2.20\pm 0.06$ & $12.7\pm 0.4$ & $8.0\pm0.2$ \\
-10	& $34.8\pm0.5$ & $7.6\pm0.1$ & $2.31\pm 0.06$ & $26.4\pm 0.6$ & $10.5\pm0.2$ \\
-25	& $54.5\pm1.7$ & $7.05\pm0.2$ & $2.8\pm 0.22$ & $46.5\pm 1.9$ & $16.5\pm0.5$\\
\hline
\multicolumn{6}{|c|}{SICCAS - CMS} \\ \hline
20 & $14.1\pm0.5$ & $8.0\pm0.1$ & $1.71\pm 0.06$ & $5.3\pm 0.3$ & $5.8\pm0.2$ \\
5 & $20.7\pm0.5$ & $7.8\pm0.1$ & $2.0\pm 0.06$ & $12.1\pm 0.4$ & $6.9\pm0.2$ \\
-10	& $31.7\pm0.5$ & $7.2\pm0.1$ & $2.33\pm 0.06$  & $23.7\pm 0.6$ & $9.8\pm0.2$ \\
-25	& $51.5\pm1.7$ & $6.5\pm0.2$ & $2.6\pm 0.22$ & $44\pm 1.9$ & $15.9\pm0.5$\\
\hline
\multicolumn{6}{|c|}{SICCAS - Y Doped} \\ \hline
20 & $15.0\pm0.5$ & $8.75\pm0.1$ & $1.67\pm 0.06$ & $5.4\pm 0.3$ & $6.6\pm0.2$ \\
5 & $22.2\pm0.5$ & $9.7\pm0.1$ & $2.06\pm 0.06$ & $11.65\pm 0.4$ & $7.9\pm0.2$ \\
-10	& $33.0\pm0.5$ & $8.8\pm0.1$ & $2.37\pm 0.06$  & $23.4\pm 0.6$ & $10.2\pm0.2$ \\
-25	& $53.5\pm1.7$ & $7.5\pm0.2$ & $2.65\pm 0.22$ & $45.5\pm 1.9$ & $15.5\pm0.5$\\
\hline
\hline
\multicolumn{6}{|c|}{Systematic uncertainities - All doped Crystals  } \\ \hline
20 & $\pm 0.8$ & $\pm 0.55$ & $\pm 0.1$ & $\pm 0.9$ & $\pm 0.1$ \\
5 & $\pm 1.1$ & $\pm 0.55$ & $\pm 0.1$ & $\pm 1.2$ & $\pm 0.1$ \\
-10	& $\pm 1.7$ & $\pm 0.5$ & $\pm 0.2$ & $\pm 1.7$ & $\pm 0.1$ \\
-25	& $\pm 2.7$ & $\pm 0.5$ & $\pm 0.2$ & $\pm 2.2$ & $\pm 0.1$ \\
\hline
\end{tabular}
\end{table}

\subsection{Enhanced analysis methods} \label{sec:Enhanced Analysis}
\paragraph{}
30000 events were acquired for each crystal (EPIC, SICCAS CMS, SICCAS-Ydoped, and CRYTUR-Panda II) and at each temperature (20 $\celsius$, 5 $\celsius$, -10 $\celsius$ and -25 $\celsius$). From Mont\'e-Carlo, we get that the mean energy deposited in the PbWO$_{4}$ crystals amount to 432 keV.

We noticed that, unlike in the simulation study above, when fitting measured data, the Cherenkov yield tend to be adjusted at values significantly lower than what we know from the Mont\'e Carlo simulation. Thus, since the Cherenkov light yield can be computed and does not depend on temperature, we set its value in the fitting algorithm to its theoretical value.

As introduced in section \ref{sec:SignalProcess} and suggested in section \ref{sec:PbWO4_FastLum}, we found that in order to fit to the measured averaged crystal pulse shapes, the scintillation model required for doped crystal two decreasing exponentials, when for undoped crystals, three decaying exponentials and a Dirac pulse are needed. Thus we used the same scintillation model equation for all crystals, but for the doped crystals, the very slow component scintillation yield and time constant are set to zero in the fitting algorithm.

\begin{figure}[htbp] 
\centering % \begin{center}/\end{center} takes some additional vertical space
\includegraphics[scale= 0.55] {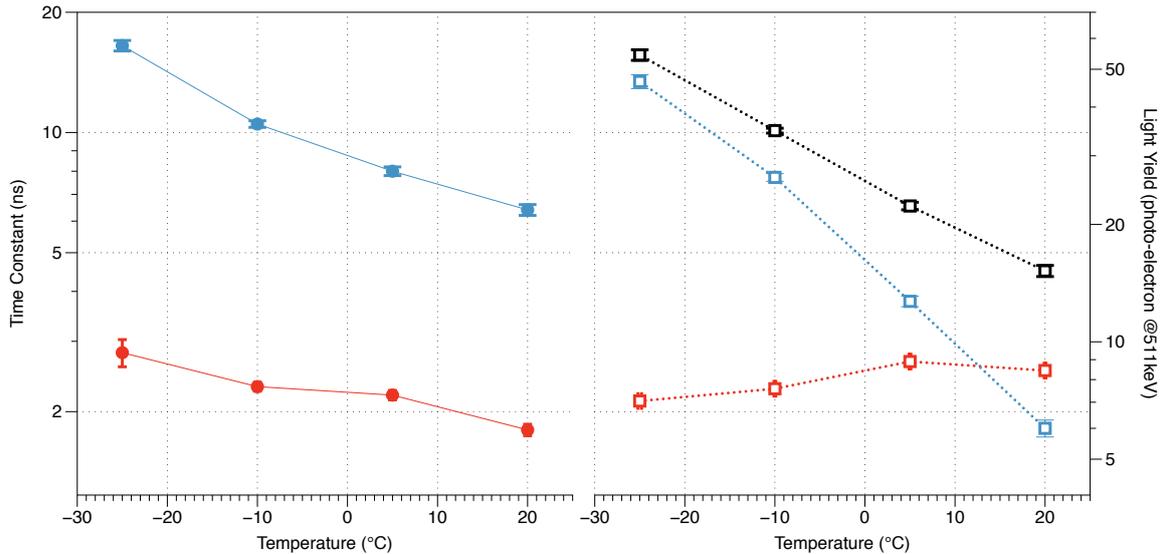}
\caption{ \label{fig:CryturResults} Measured luminescence properties on Crytur PbWO$_{4}$ crystal Sample. The fast component is shown in red, the slow component is shown in blue and the total flux is shown in black. Error bars shown account for statistical errors only. The SICCAS doped crystals show very similar properties.}
\end{figure}

\subsubsection{Taking care of post-pulse overshooting} \label{sec:Pulse Overshoot}
\paragraph{}
The integral of the averaged signal shape encodes the number of detected photo-electron and the shape of the pulse. Some additional care has to be taken in the analysis: 

First it turns out that the HT-positive PMT biasing system high-pass the pulse shape. This induces a small but significant overshoot of 2.1 \% of max amplitude that can hardly be noticed on the single photoelectron pulse shape and is relaxing with a time constant of 500 ns. When integrated on tens of nanoseconds (time needed to integrate the slow scintillation component, this overshoot subtract a significant part of the computed charge and leads to the dependence of the measured charge from the integration window. This effect can be modelled and compensate for, if we know the crystal scintillation properties.

As explained, the mean Cherenkov photo-electron yield is needed as an input in order to compute accurately the fast scintillation time constants. It can be computed from the Mont\'e Carlo. But our fitting software requires to input its value as a fraction of the total photoelectron yield. Thus we need to know the total photo-electron yield, before hand.

Thus we decided to process the data in two steps.
As a first step, the charge of the mean crystal signals pulses, integrated over 50 ns are divided by the single photo-electron charge. This provides a first raw measurement of the photo-electron yield, that is used to compute the raw mean photo-electrons number, Cherenkov light yield fraction, and through fitting, first values for the crystal scintillation properties. 
The raw scintillation time constants and yields are then used to compute the correction factors one needs to apply on light yield versus the integration time window, and thus to assess the intrinsic photo-electron yield of the crystal. We checked that after correction, calculated photo-electron yields do not depends significantly on the integration window anymore.
As a second step we use this enhanced photoelectron yield to compute the updated Cherenkov fraction, and then fit the refined crystal scintillation properties.

\subsection{Photo-electron yields and scintillation time constants} \label{sec:PE-Yield}
\begin{figure}[htbp] 
\centering % \begin{center}/\end{center} takes some additional vertical space
\includegraphics[scale= 0.60] {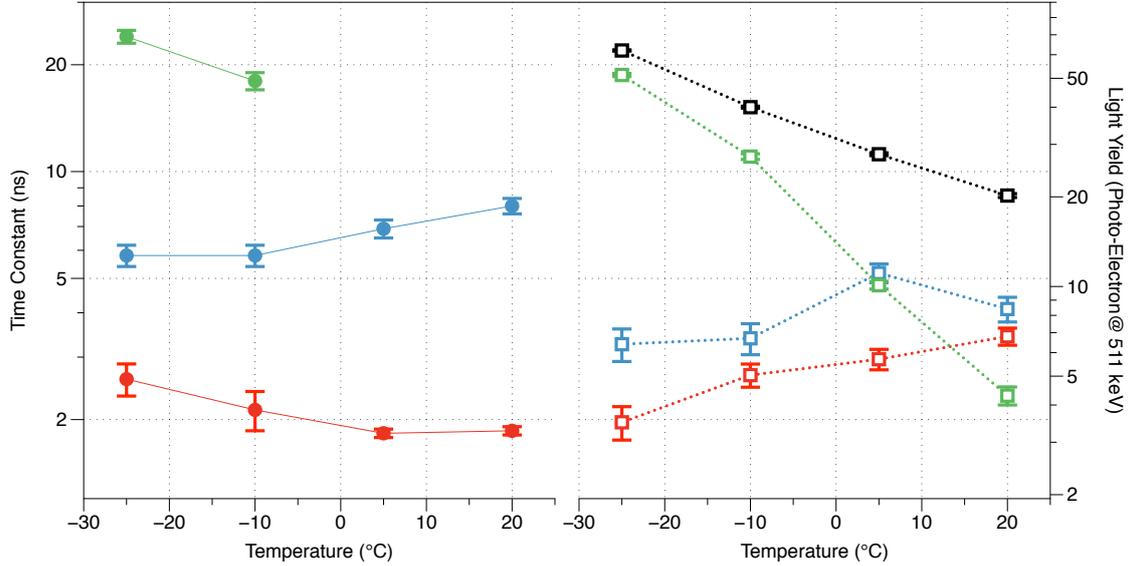}
\caption{ \label{fig:EpicResults} Measured luminescence properties on EPIC undoped PbWO$_{4}$ Crystal Sample. The fast component is shown in red, the slow component is shown in blue, and the very slow component is shown in green. The total flux is shown in black. Error bars shown account for statistical errors only. }
\end{figure}

\paragraph{}
Result are shown at tables \ref{tab:Doped_PbWO4_Results} and \ref{tab:Undoped_PbWO4_Results} and figure \ref{fig:CryturResults} and \ref{fig:EpicResults}.
We notice that the three different doping technology lead to very similar perfomances. As documented in literature, for all technologies, the slow component photo-electron yield and time constant increase a lot when temperature lowers.
This is not true for the fast component. Quite surprisingly, the fast photo-electron yields do not depend much on temperature, unlike the time constant that moves from ~1.8 ns at 20 \celsius to ~2.8 ns at -25 \celsius. Looking carefully, SICCAS-Y crystal and Crytur-PandaII display a slightly enhanced yield compared to SICCAS-CMS.

\begin{table}[htbp]
\centering
\caption{\label{tab:Undoped_PbWO4_Results} Measured luminescence properties of undoped PbWO$_4$ scintillator crystals on our apparatus for a mean energy deposited of 432 keV. $Y_{xx}$ stands for photo-electron yields, $\tau_{XX}$, for scintillation time constants. The Cherenkov contribution to yield, 0.80 photo-electron, is included in the total luminescence Yield.
Second part : Computed values of systematic errors on measurements (see paragraph \ref{sec:Systematics}) 
}

\smallskip
\begin{tabular}{|c|c|c|c|c|c|c|c|}
\hline
Temp. & $Y_{Total}$ & $Y_{Fast}$ & $\tau_{Fast}$ & $Y_{Slow}$ & $\tau_{Slow}$ & $Y_{VSlow}$ & $\tau_{VSlow}$ \\
(\celsius) & (PE) & (PE) & (ns) & (PE) & (ns) & (PE) & (ns) \\
\hline
\multicolumn{8}{|c|}{EPIC (undoped) crystal} \\ \hline
20 & $20.2\pm 0.3$ & $6.8\pm0.45$ & $1.86\pm 0.05$ & $8.4\pm 0.8$ & $8.0\pm0.4$ & $4.3\pm0.3$ & -- \\
5 & $27.8\pm 0.3$ & $5.7\pm0.45$ & $1.83\pm0.05$ & $11.1\pm 0.8$ & $6.9\pm0.4$ & $10.1\pm0.3$ & -- \\
-10	& $40\pm 0.3$ & $5.0\pm0.45$ & $2.1\pm 0.27$  & $6.7\pm 0.8$ & $5.8\pm0.4$ & $27.3\pm0.6$ & $18\pm 1$ \\
-25	& $62\pm 0.3$ & $3.5\pm0.45$ & $2.6\pm 0.27$ & $6.4\pm 0.8$ & $5.8\pm0.4$ & $51.3\pm0.6$ & $24\pm 1$ \\
\hline
\hline
\multicolumn{8}{|c|}{Systematics uncertainities} \\ \hline
20 & $\pm 1.2$ & $\pm 0.5$ & $\pm 0.1$ & $\pm 0.8$ & $\pm 0.3$ & $\pm 0.45$ & -- \\
5 & $\pm 1.6$ & $\pm 0.5$ & $\pm 0.1$ & $\pm 0.9$ & $\pm 0.3$ & $\pm 0.65$ & -- \\
-10	& $\pm 2.1$ & $\pm 0.5$ & $\pm 0.1$ & $\pm 0.8$ & $\pm 0.3$ & $\pm 1.4$ & $\pm 1.0$ \\
-25	& $\pm 3.2$ & $\pm 0.45$ & $\pm 0.1$ & $\pm 0.8$ & $\pm 0.3$ & $\pm 2.6$ & $\pm 1.0$ \\
\hline
\end{tabular}
\end{table}

The only undoped crystal shows a significantly higher photo-electron yield. This luminescence is shared in three components. The fast component is of lower yield compared to doped crystal, with shorter time constants. The slow component shows a time constant close those measured on doped crystals, but of reduced yield at low temperature. The very slow time constant contributes to most of the additional yield and increases quickly with temperature. Our apparatus focussed on quantifying fast scintillation components. Thus at room temperature where the yield of very slow component is small and spread on a large time window, the measurement of the time constant is very inaccurate, and thus not shown.

\section{ Uncertainties on Photo-electron yields and scintillation time constants}\label{sec:Systematics}
We considered the following sources of error:

\subsection{Mont\'e Carlo simulation}
\paragraph{}
	The Mont\'e Carlo simulation is used in the analysis twice.
	
		First it allows to compute the light collection time distribution in the BRS detector. As explained in the paragraph \ref{sec:SimuOptSurf} measuring the crystal surface allowed us to compute models of the optical photon diffusion at the crystal surfaces. In the following analysis, choosing a light collection time distribution based on any of the measured data does not impacts the fitted scintillation parameters but, at the lowest temperatures, on the fast scintillation parameters and is taken into account.
		
		Then we used the simulation to compute the Cherenkov photo-electron yield. Assuming a 20\% error, after processing, this induce an error 0.16 photo-electron on the fast scintillation yield, and negligible error on other fast or slow scintillation recovered parameters.

%\begin{figure}[htbp] 
%\centering % \begin{center}/\end{center} takes some additional vertical space
%\includegraphics[scale= 0.47] {Figures/EPIC_residus_SPE1_SPE3_HT.png}
%\caption{\label{fig:ResAfterNoisSub}  Timeline of the difference between the measured and the fitted pulse shape.
%\emph{Left:} We used the optimised SPE pulse shape in the measurements processing. 
%\emph{Right:} We subtracted our noise estimation from the aveeraged SPE pulse shape before processing. Subtraction of the noise estimation does not improve the residuals of the fit. See text for more details.
%}
%\end{figure}

\subsection{BRS photo-multiplier calibration}
\paragraph{} 
	This includes two different sources of uncertainity.
	
	The first is on the single photo-electron charge measurement. The apparatus electronics requires a stabilization/warmup time quite long: 45 minutes was found to be safe. Then we measured a single photo-electron mean charge value of $(1.6 \pm0.08)$ $10^{7}$. This 5\% error impacts all the measured scintillation yields values.
		
	The second is the single photo-electron mean shape measurement.
	We know from the timeline of the difference between the fitted and measured pulse that our pulse shape model is still not optimal: the typical $\chi^2$ after fitting is 3000 for 219 degree of freedom. This is due to the remaining laser-induced noise on the PMT readout, during the BRS photomultiplier calibration. Work presented at paragraph \ref{sec:SPE MeanShape} allowed us to damp the noise on the modeled pulse shape, but in the main pulse part, that was left unfiltered so as not to bias the analysis.
	
	%We took data, putting a black tissue in front of the optical fiber output and triggering on the Pilas laser synchronization signal. Unfortunately, due to our lab contingencies (move up), our apparatus was not the same as for measurements and cannot be reproduced today. Averaging these noise-only data, we measured a well structured noise timeline. After normalisation, we substracted the noise timeline from the raw mean shape  and processed it into an hopefully "optimized pulse shape, noise substracted". Using this modified pulse shape in the processing of our measurements lead to difference timeline shown at figure \ref{fig:ResAfterNoisSub}. 
	%We notice that instead of reducing the amplitude of the residuals, these have increased. Indeed we somehow overcompensated the error timeline in the main pulse time window by a factor of four, the later part of the residual staying mostly the same. We decided nevertheless to run our analysis using also this modified pulse shape in order to assess the systematic error due to imperfect BRS Pulse Shape.
	We tried to modify the SPE mean shape, in order to minimize the residuals after full reprocessing and fitting our data. We did not succeeded to cancel the residuals, but we were able to put an upper value on the systematic uncertainty on our measurements.

\subsection{Crystal temperature and reproducibility}
\paragraph{}
	During the ramp up/down of the crystal temperature, we waited two hours for the crystal temperature to stabilize, before taking measurements. We evaluated this error by taking two measurements, one during the ramp down temperature, one during the ramp up run. This also allows us to evaluate the reproducibility of our measurements.

\subsection{The others factors}
\paragraph{}
	Photo-multiplier dark count. The dark count rate on our photomultiplier, biased with a positive high voltage has been measured below 100 Hz. This induce a probability to add a photoelectron within a time window of 300 ns of 3.10$^{-5}$, thus negligible.

	Pilas laser, YAP spectrometer and Wavecatcher time resolution. These uncertanies are dominated by the YAP spectrometer time resolution uncertanity $104 \pm 4.7$ ps (1 standard deviation) at 511 keV, thus negligible.
	
	- The $^{22}$Na radioactive source emits a 1.27 MeV gamma ray in coincidence with the 
	$\beta+$ decay in a solid angle of 4$\pi$ Steradian. The probability of this photon to interact in the PbWO$_{4}$ crystal is computed to 0,2\%, thus does not impact our light yield measurement significantly.
			% Calcul des probabilités d'interaction des photons de 511 keV et de 1.27 MeV dans le cristal de PWO 
			% Seff(511 keV) PWO = 1,37e-1 cm2/g, Seff(1.27 MeV) PWO = 5,67e-2 cm2/g
			% ASolid(BRS)/4pi = 0.0057
			% Pi(1.27 MeV) = (1- exp(-Seff*8,28*1))*ASolid(BRS)/4pi,= 0,213 % 
			
	We evaluated each of these systematics and when relevant summed their contribution quadraticaly. Three main contributions drives the systematic uncertainty budget. The BRS SPE charge calibration drives most light yield uncertainties. The remaining noise on SPE pulse shape impacts the time constant measurements and the remaining uncertainty on LUT impact the fast components measurements.
	The results are shown a tables \ref{tab:Doped_PbWO4_Results} and \ref{tab:Undoped_PbWO4_Results}. Systematic errors dominate the error budget on most measured values.
	
\begin{table}[htbp]
\centering
\caption{\label{tab:Scint_Yield} Estimation of PbWO$_4$ scintillation yield versus temperature}
\smallskip
\begin{tabular}{|c|c|c|c|c|}
\hline
Temp. (\celsius) & 20 & 5 & -10 & -25 \\
\hline
\multicolumn{5}{|c|} {All doped crystals} \\ \hline
Fast+Slow Yield ($\gamma$/MeV) & 330 & 500 & 790 & 1250 \\
\hline
\multicolumn{5}{|c|}{EPIC crystals} \\ \hline
Fast+Slow Yield ($\gamma$/MeV)& 350 & 390 & 270 & 230\\
Very Slow Yield ($\gamma$/MeV)& 130 & 310 & 850 & 1600\\
\hline
\end{tabular}
\end{table}

\section{Scintillation properties}
\paragraph{}
	After measuring the photo-electron light yield of the PbWO$_{4}$  crystal on the R11265-100 photomultiplier, we considered computing the photon light yield in the three PbWO$_{4}$  scintillating wavelengths. An accurate work would require to know the PMT photocathode efficiency curve and the scintillation spectrum of the two or three scintillation bands for each of the crystal doping technology.
	To date we do not have in lab the tools to measure the PMT's photocathode efficiency curve. The literature shows very significant discrepancies between suppliers and devices, related to photocathode thickness and manufacturers' know-how. Thus the following numbers \emph{should be taken as orders of magnitude} and we won't try to quantify errors.
	
	From the Mont\'e Carlo simulation data we compute, for optical photons produced within the PbWO$_4$ crystal in our test setup, the light collection efficiency of the BRS PMT photocathode: 33\%.
From the \citep{Motta2005a, Shao_2001, Korzhik_2003} we computed the quantum efficiency of a bialkali photocathode averaged over the PbWO$_4$:Y scintillation spectrum, 26\% and of the undoped crystal slow component PbWO$_4$ peaking at 500 nm, 19\%. From these values and the measured light yields, we estimated the scintillation yield of the fast + slow scintillation components and of the very slow component versus temperature, shown at table \ref{tab:Scint_Yield}.

	It is worth noticing that would a crystal scintillate at red wavelengths, an/or $\mu$s time constant, our apparatus would not be able to measure this component. The BRS PMT's bialkali photo-electric layer is efficient at blue and green wavelengths only, and $\mu$s scintillation time constants would not show significant flux on a 50 ns time window.
	
\section{Discussion}
\paragraph{}
	Measurements of PbWO$_4$ scintillating properties from gamma ray interaction in the crystal turned out to be a difficult task. Many instrumental effects have to be deconvoluted from the measurement in order to recover the scintillating properties. 
	
	Our measurements reproduce the main qualitative features of PbWO$_4$ crystal quoted in the literature. Quantitatively though, we measured significantly shorter time constant and larger light yield than those published in the reference articles of the field \cite{Annenkov2000, Kobayashi_1999}.
	
	The shorter measured scintillation time constants are easily understood because we used thin scintillating crystals and we subtracted the light collection time distribution of our apparatus, which was not usually done in previous works.
	The larger scintillation yields are easily understood because in this measurement we optimized the light collection efficiency by using small grounded crystal, with a large coupling crystal/PMT surface. In addition, we took into account this light collection efficiency of our measurement, when computing photon light yield.  
	
	The R11265-100 PMT, at the time we initiated this work, was one of the best \emph{non-MCP} PMT technology as far as TTS is concerned (specified to 270 ps FWHM, rise time 1.3 ns). For the shorter scintillation components, we considered the use of MCP-PMT. These technologies allow much shorter pulse shapes, that should ease the deconvolution processing, and increase sensitivity to short time constants. But when we calibrated their Single Photo-Electron response these devices showed in the time distribution, in addition to the main peak, a long tail (ns scale, 25\% of statistics \citep{Canot2019Dec}). This tail would need to be deconvoluted from measurements, increasing the analysis complexity. MCP also often induce a photo-electron collection inefficiency, up to 40\%, significant when one want to measure the properties of low light yield scintillating crystals. Finally today one inch-square MCP-PMT are still expensive. Thus for these measurements the  R11265-100 appeared to be the best option.

\subsection{Prospects for the Light Yield improvement.}
\paragraph{}
An improvement of the PbWO$_4$ light yield by various doping seems to have exhausted itself. Different luminescence centers compete to regular WO$_{4}^{2-}$ groups to catch nonequilibrium carriers, so they dont get so many electrons. Due to this reason no impressive progress to improve light yield of the scintillator with various doping has been obtained so far \cite{Annenkov2000}. Mo doping of PbWO$_4$ do can triple the measured light yield. Then the maximum of the luminescence is shifted to 500 nm. However a large fraction of the scintillation kinetics is now in a slow component with decay constant of the order of 300 ns \cite{Bohm1998}.

For the ClearMind project a prospective way will be to search for the the optimal combination of operational temperature and concentration of a few dopings, which might look unusual from the point of view of the previous development of the material for High Energy physics. Here, to obtain better statistics in a shorter time, we plan to focus on an increase of the amount of the photons in the leading edge of the scintillation pulse, as a complement to the collection of the Cherenkov light in the crystal.

\acknowledgments
We want to thank Carlotta Trigila and Emilie Roncali (UC. Davis) for their support using the Davis LUT model and associated softwares.
We are grateful for the support and seed funding from the CEA, Programme Exploratoire Bottom-Up, under grant No. 17P103-CLEAR-MIND and from the French National Research Agency under grant No. ANR-19-CE19-0009-01.

% We suggest to always provide author, title and journal data:
% in short all the informations that clearly identify a document.

\bibliographystyle{JHEP}
\bibliography{Scintillators,PET,Detectors,Softwares} {}

%\begin{thebibliography}{99}

%\bibitem{a}
%Author, \emph{Title}, \emph{J. Abbrev.} {\bf vol} (year) pg.

%\bibitem{b}
%Author, \emph{Title},
%arxiv:1234.5678.

%\bibitem{c}
%Author, \emph{Title},
%Publisher (year).

% Please avoid comments such as "For a review'', "For some examples",
% "and references therein" or move them in the text. In general,
% please leave only references in the bibliography and move all
% accessory text in footnotes.
% Also, please have only one work for each \bibitem.
%\end{thebibliography}

\end{document}